\documentclass[twocolumn,aps,pra,superscriptaddress,showpacs,preprintnumbers]{revtex4-1}%
\usepackage[latin9]{inputenc}
\usepackage{units}
\usepackage{amsmath}
\usepackage{graphicx}
\usepackage{amssymb}
\usepackage{amsfonts}
\usepackage{verbatim}
\usepackage{natbib}
\usepackage[english]{babel}
\usepackage[dvipdfm]{hyperref}

\begin{document}

\title{Evanescent straight tapered-fiber coupling of ultra-high Q optomechanical micro-resonators in a low-vibration helium-4 exchange-gas cryostat}

\author{R.~Rivi\`{e}re}
\affiliation{\'{E}cole Polytechnique F\'{e}d\'{e}rale de Lausanne (EPFL), 1015 Lausanne, Switzerland}
\affiliation{Max-Planck-Institut f{\"u}r Quantenoptik, 85748 Garching, Germany}
\author{O.~Arcizet}
\affiliation{Institut N\'eel/CNRS and Universit\'{e} Joseph Fourier, 38042 Grenoble, France}
\author{A.~Schliesser}
\affiliation{\'{E}cole Polytechnique F\'{e}d\'{e}rale de Lausanne (EPFL), 1015 Lausanne, Switzerland}
\affiliation{Max-Planck-Institut f{\"u}r Quantenoptik, 85748 Garching, Germany}
\author{T.~J.~Kippenberg}
\email{tobias.kippenberg@epfl.ch}
\affiliation{\'{E}cole Polytechnique F\'{e}d\'{e}rale de Lausanne (EPFL), 1015 Lausanne, Switzerland}
\affiliation{Max-Planck-Institut f{\"u}r Quantenoptik, 85748 Garching, Germany}

\begin{abstract}
We developed an apparatus to couple a 50-$\mu$m diameter whispering-gallery silica microtoroidal resonator in a helium-4 cryostat using a straight optical tapered-fiber at 1550\,nm wavelength.
On a top-loading probe specifically adapted for increased mechanical stability, we use a specifically-developed ``cryotaper'' to optically probe the cavity, allowing thus to record the calibrated mecha\-nical spectrum of the optomechanical system at low temperatures.
We then demonstrate excellent thermalization of a 63-MHz mechanical mode of a toroidal resonator down to the cryostat's base temperature of 1.65\,K, thereby proving the viability of the cryogenic refrigeration via heat conduction through static low-pressure exchange gas.
In the context of optomechanics, we therefore provide a versatile and powerful tool with state-of-the-art performances in optical coupling efficiency, mechanical stability and cryogenic cooling.
\end{abstract}

\maketitle

\affiliation{Max Planck Institut f{\"u}r Quantenoptik, 85748 Garching, Germany}
\affiliation{\'{E}cole Polytechnique F\'{e}d\'{e}rale de Lausanne (EPFL), 1015 Lausanne, Switzerland}

\section{Introduction}

Cavity optomechanics has attracted considerable attention recently, to become now a major area in mesoscopic physics \cite{Kippenberg2007,Kippenberg2008,Marquardt2009,Favero2009,Aspelmeyer2010}.
By exploiting the coupling of an optical mode and a mechanical mode, the measurement of quantum effects involving a macroscopic mechanical structure is now possible.
Very recently, using the dynamical backaction of light, macroscopic mechanical oscillators have even been prepared in low entropy states, very close to the quantum ground state \cite{Verhagen2011a,Chan2011,Teufel2011}.
The study of the quantum manifestations of the optomechanical interaction is however so far only possible if the thermal fluctuations resulting from the coupling of the mechanical oscillator with the environment are reduced beforehand.
To do this, the optomechanical systems are placed in cryostats based on the cooling technology provided by the evaporation of $^4$He \cite{Schliesser2009a,Park2009,Groeblacher2009} or $^3$He \cite{Riviere2011}. 
In our case, the experimental structure under study is a silica whispering-gallery microresonator hosting high quality factor ($\sim 10^8$) optical modes in the visible and near infrared and mechanical modes with resonance frequencies exceeding 50\,MHz \cite{Kippenberg2005,Schliesser2006}.

The implementation of the experiment in a cryogenic environment raises two major problems whose resolution is described in this article.
First, the optical cavity has to be coupled in a stable way to an optical coherent field.
For our experiment, we have selected the tapered-fiber coupling technique for its superiority in terms of efficiency compared to other possible techniques \cite{Spillane2003}.
A so-called ``cryo\-taper'' is formed by attaching the straight tapered-fiber to an adequate glass fiber-holder.
It is placed on a dedicated mechanical construction, therefore enabling a proper optical coupling by retaining the taper's tension and integrity, and by insuring a stable toroid-taper gap ($< 1\,\mu$m) down to low temperatures (1.65\,K).
Second, the cryogenic system has to extract the heat in the sample generated by all sources, including by the absorption of a high input power required in coo\-ling experiments ($\mathcal{O}(0.1)\,$mW) \cite{Schliesser2009a,Riviere2011} confined in a small modal volume ($\mathcal{O}(100)\,\mu$m$^3$) \cite{Kippenberg2004a}.
This heat source is moreover located in a weakly thermally-conducting structure made of silica \cite{Anetsberger2008}.
Therefore, the experiment is conducted in a $^4$He exchange-gas cryostat described here that enables complete thermalization of the entire system (at low optical powers), owing to the isotropic thermal conduction provided by the exchange gas.

We present in this article the technical details of the experimental arrangement using the above-mentioned no\-velties, which has been used since then in a number of previously reported experiments \cite{Schliesser2009a,Weis2010,Riviere2011,Verhagen2011a}.

\section{The low-pressure helium-4 exchange-gas cryostat}

\begin{figure}[t!]
  \centering
  \includegraphics[width=\columnwidth]{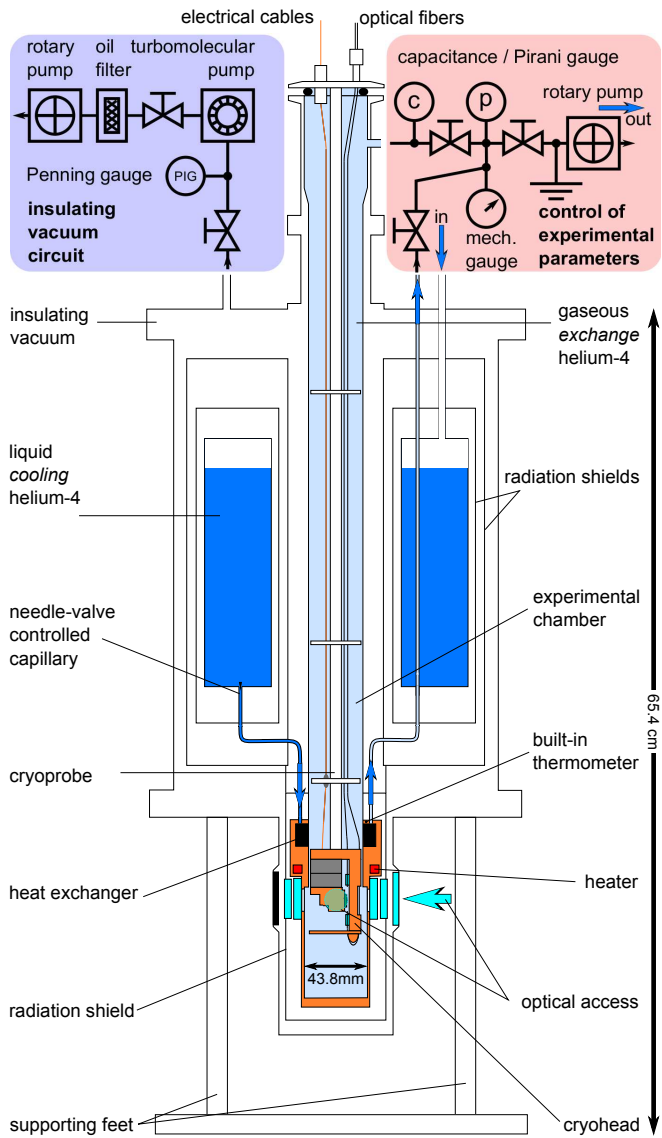}
  \caption{Technical layout of the helium-4 cryostat descri\-bing the commercial Oxford Instrument OptistatSXM static exchange gas cryostat (to scale), the modified top-loading cryo\-probe supporting the cryogenic head, the optical fiber and electrical cable layout, the optical access and the schematic vacuum circuits.}
  \label{figure1}
\end{figure}

The thermalization of a 4.5-$\mu$m diameter GaAs/AlGaAs whispering-gallery mode resonator to the base temperature of a $^4$He continuous-flow cryostat is known to be a technical challenge (See Ref.~\cite{Srinivasan2007}).
This difficulty is attributed to the inefficient extraction mechanism of the heat from the disk-shaped sample placed in high vacuum, relying exclusively on thermal conduction through the supporting structure attached to a ``cold finger''.
Even by enlarging the thermal anchoring to the cooled part of the cryostat using copper braids, the effective cooling power on the sample is too small to ensure complete thermalization, inducing therefore large temperature gradients across the sample.
Although our toroid-shaped sample is made of SiO$_2$ and has a $50$-$\mu$m diameter, this first thermalization attempt inspired us to investigate a different type of refrigeration mechanism.
Consequently, we use for efficient cryogenic cooling of the sample a low-pressure \textit{exchange}-gas, itself cooled by the cryogenic device.
An exchange-gas pressure on the order of 1\,mbar is used since it has been demonstrated in Ref.~\cite{Anetsberger2008} that in this pressure range the gas does not significantly damp the mechanical mode under study.

The utilized apparatus is a low-pressure $^4$He exchange-gas cryostat based on the Oxford Instrument OptistatSXM model.
The cooling process is named \textit{static} exchange-gas cooling since the volume into which the cryo\-probe is inserted is not subjected to any gas circulation.
This class of device provides a cooling power exceeding 1\,mW at liquid $^4$He temperatures.

Figure \ref{figure1} shows an on-scale simplified drawing of the cryostat.
This device does not have a liquid N$_2$ shield as a first cooling stage.
Instead, radiation shields made of a stack of insulating layers are directly cooled by the exhaust gas from the 4.3-L liquid $^4$He tank.

From this reservoir, the \textit{cooling} liquid $^4$He is admitted into a capillary tube through an adjustable needle valve.
The coolant is then circulated using a 40\,m$^3$/h rotary pump (Oxford Instrument EPS40) and evaporated at a tunable rate.
By adjusting the admission and pumping flows, the vapor pressure is controlled, thus setting the working temperature and the cooling power of the cryostat.
The heat is extracted by the flux of \textit{cooling} helium from the experimental chamber via the heat exchanger, the \textit{cooling} helium's evaporation being forced by the rotary pump.
The large pumping power of the selected pump provides the high evaporation rate required to reach the targeted temperature.
Inside the experimental chamber, the \textit{exchange} $^4$He gas thermalizes the sample with the chamber's cooled walls.
By using a built-in electric heater and temperature sensor, the cryostat temperature is locked to $\pm 0.1$\,K accuracy over 10\,min for the temperature range of interest down to 1.65\,K.

\begin{figure}[h,floatfix]
  \centering
  \includegraphics[scale=0.85]{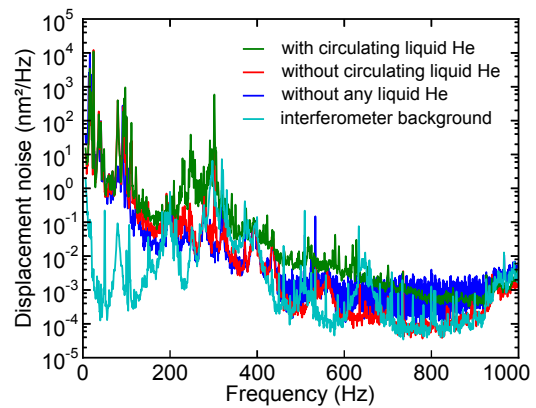}
  \caption{Vibration spectra of the cryohead for various regimes of the cryostat.
The measurements are performed with an optical interferometer fixed to the table, probing through the front access a mirror attached to the cryohead.}
  \label{figure3}
\end{figure}

The specific mechanical construction provided by the manufacturer ensures a low vibration level of the whole device, sufficient to perform sensitive experiments such as scanning tunneling microscopy \cite{Yi2004}.
In addition, because of its rather small height (1\,m including the necessary vacuum components), the cryostat can be placed on a conventional optical table and benefits from its acoustic isolation.
To cut off extra vibrations from the rotary pumps, the connecting hoses are firmly clamped to the ground.
The measurement of the acoustic vibrations of the cryohead compared to the table taken as an external reference (Fig.~\ref{figure3}) reveals however an important contribution from the circulating \textit{cooling} liquid He between 200 and 320\,Hz.
The rms displacement in this cryogenic cooling regime reaches then 150\,nm, exceeding by more than 10\,nm the value measured without circulating liquid.

Two separated vacuum circuits are connected to the setup: one for pumping the insulation vacuum prior to cool down (Fig.~\ref{figure1}, blue inset) and another one to re\-gulate the pressure of the \textit{cooling} and \textit{exchange} $^4$He gas (Fig.~\ref{figure1}, red inset).
With this configuration, no significant degradation of the cavity's optical properties (e.~g.\@ by oil diffusion from the pump) was observed.
The piping presented schematically in Fig.~\ref{figure1} allows to fill the experimental chamber with \textit{exchange} $^4$He gas directly from the evaporation port of the capillary tube, therefore providing a very pure source of gas.
This arrangement avoids the contamination of the chamber with unwanted gases such as H$_2$O or N$_2$ that may freeze on the sample and deteriorate its properties.
Eventually, frozen contaminants can be sublimated by pointing a 10-W Coherent Verdi laser (532\,nm wavelength) directly onto the sample through the front optical window of the cryostat.

\section{The cryohead}

\begin{figure}[t!]
  \centering
  \includegraphics[width=\columnwidth]{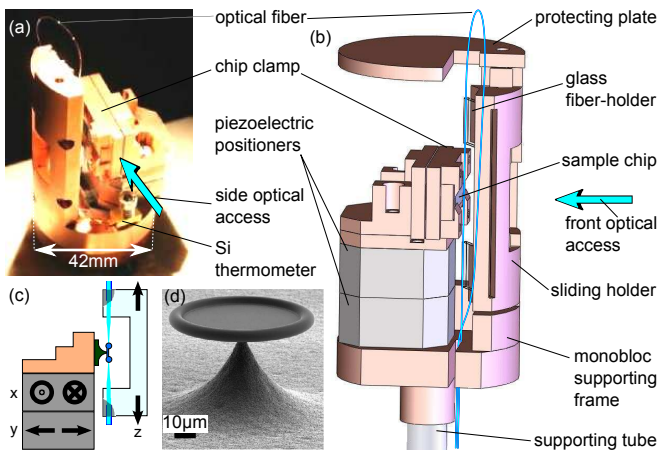}
  \caption{Photograph and technical layout of the cryohead.
(a) Photograph of the cryohead without the protecting plate.
The thermometer is a Lakeshore DT-670B-CO Si diode.
(b) 3D rendering of the cryohead showing the piezoelectric positioners (Attocube Systems ANPx101/LT/HV, ``low temperature high vacuum'' version) supporting the 3-part sample holder and the cryotaper placed on the mechanical slide, held by the rigid frame.
The protecting plate avoids damaging shocks during the cryoprobe insertion.
(c) Off-scale symbolic drawing of the coupling mechanism.
The two piezoelectric positioners displace the clamped chip in the x-y plane so as to approach the desired toroid in the near-field of the mechanical-slide supported cryotaper when the cryoprobe is inserted and cooled down in the cryostat.
The z-position is manually adjusted by sliding the holder prior to cooling.
(d) Scanning-electron micrograph of a typical silica microtoroidal resonator.
}
  \label{figure2}
\end{figure}

To couple the optical mode of the microcavity, a single-mode tapered-fiber of diameter on the order of the optical wavelength $\lambda$ (1550\,nm) is approached in the near-field of the whispering-gallery mode.
This evanescent part of the optical field radially extends outside the cavity over a distance decreasing approximatively with $e^{-2 \pi \sqrt{n^2-1} (r-R) / \lambda}$, where $n$ is the refractive index of the cavity's silica, $R$ the cavity radius and $r$ the radial coordinate \cite{Anetsberger2011}.
Simultaneously, contact of the resonator with the tapered fiber has to be avoided in order not to deteriorate the mechanical properties of the device.
The coupling therefore necessitates a precision of the spatial placement of the fiber much finer than $\lambda$.
To this end, we developed a particularly stable mechanical construction with a positioning system based on dedicated commercial stages exploiting both slip-stick (coarse approach) and slow (fine approach) motion of piezoelectric elements.
The whole setup is placed at the cold point of the experimental chamber of the cryostat using a top-loading probe.

The cryoprobe shown in Fig.~\ref{figure1} is a probe from the manufacturer modified to host the cryohead comprising the whole coupling setup (Fig.~\ref{figure2}).
This head is compacted to fit into the experimental chamber of diameter $43.8\,$mm.
To avoid that the elements of the head vibrate under acoustic excitation from the environment and jeopardize the coupling stability, the cryohead is specifically designed in a compact way, avoiding floppy mechanical elements of low resonance frequencies and high oscillation amplitudes.
The present design proves to be outstandingly stable.
Coupling can then be successfully performed, even in a noisy laboratory environment and in the presence of the circulating \textit{cooling} helium that induces vibrations, when a constant temperature is reached.
Furthermore, \textit{exchange} gas displacement that may be induced for instance by unwanted local convection does not have any recordable effect on the coupling stability.

Figures \ref{figure2} (a) and (b) show a photograph and a 3D rendering of the cryohead, respectively.
The sample is clamped using a mechanical claw onto a holder that can be smoothly slided and attached to the piezoelectric displacers without risks for the nearby cryotaper (detailed in the next section).
Opposite to the sample chip, the cryotaper is attached to a mechanical slide.
Prior to cooling, this construction allows scanning longitudinally the tapered region until finding the appropriate zone of the gradual taper transition allowing proper phase matching \cite{Knight1997}.
Under normal experimental conditions, it does not need further adjustment at low temperatures.
Figure \ref{figure2} (c) symbolically represents the translation axis allowed on the cryohead to ensure optimum coupling to a typical microtoroidal resonator shown in Fig.~\ref{figure2} (d).

\section{The cryotaper}

\begin{figure*}[t]
  \centering
  \includegraphics[scale=0.85]{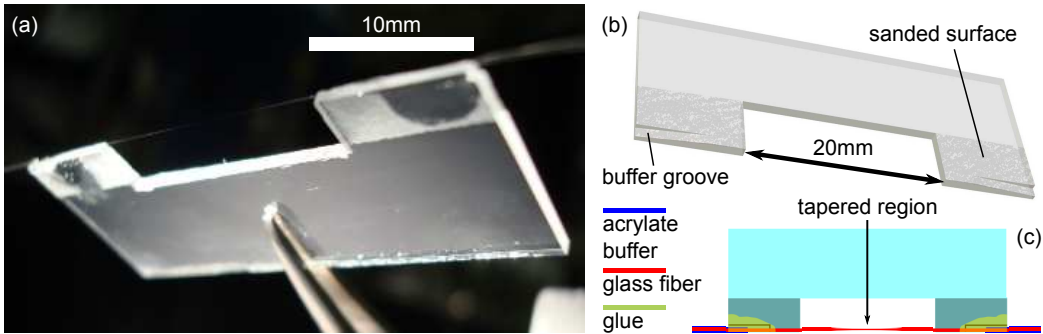}
  \caption{Photograph and technical layout of the cryotaper.
(a) Photograph of a finalized cryotaper.
(b) On-scale 3D rendering of the cryotaper showing the groove for hosting the buffer of the optical fiber (245\,$\mu$m diameter) and the sanded surface for better glue adhesion.
(c) Technical layout describing the usual configuration of the cryotaper.
To avoid shearing of the glass fiber during manipulation, the taper is crafted to dimensions such that the acrylate buffer is glued to the glass slide with simultaneously the uncovered central part of the glass fiber being still in contact with the support to avoid having a long suspended length subjected to large amplitude vibrations.
}
  \label{figure4}
\end{figure*}

The most delicate part of the cryohead is the cryo\-taper, because of the extreme fragility of the tapered part of micrometer-size diameter.
Although other coupling techniques using free-space beams \cite{Park2009}, prisms \cite{Gorodetsky1994}, eroded \cite{Laine1999} or angle-cleaved optical fibers \cite{Ilchenko1999} are mechanically less constraining, the tapered-fiber coupling is employed for its higher efficiency \cite{Spillane2003}.
Additionally, it enables to easily reach the overcoupled regime by displa\-cing the fiber in the azimuthal plane of the cavity mode.
These two properties make the tapered fiber an optimal coupling system for sensitive optical detections.

Unlike other groups using a bent tapered fiber \cite{Srinivasan2007}, we have chosen to keep the tapered fiber straight to ensure a maximal mechanical stability and adequate tensioning.
To keep it straightly tensioned after cool down, the tapered fiber is glued to a C-shaped piece of glass presented in Fig.~\ref{figure4} (a) and (b).
Since both glued parts are made of essentially the same material, the tapered fiber remains properly tensioned at all temperatures as experimentally verified.
A UV epoxy glue is specifically used because it does not contract when cured (it can also sustain several temperature cycles).
Thus, the fiber tension properly adjusted during the fabrication process remains unchanged during the gluing and thereafter du\-ring the cool down operation.
Indeed, typically, at all the temperatures the toroid can be properly coupled with a tapered fiber that neither breaks nor distends, as visually monitored through the optical accesses.
In Fig.~\ref{figure4} (c), the geometrical arrangement of the cryotaper is schematized, showing the groove in the glass slide hosting the acrylate buffer of the glass fiber and the sanded surface ensuring proper glue adhesion.

\begin{figure*}
  \centering
  \includegraphics[scale=0.85]{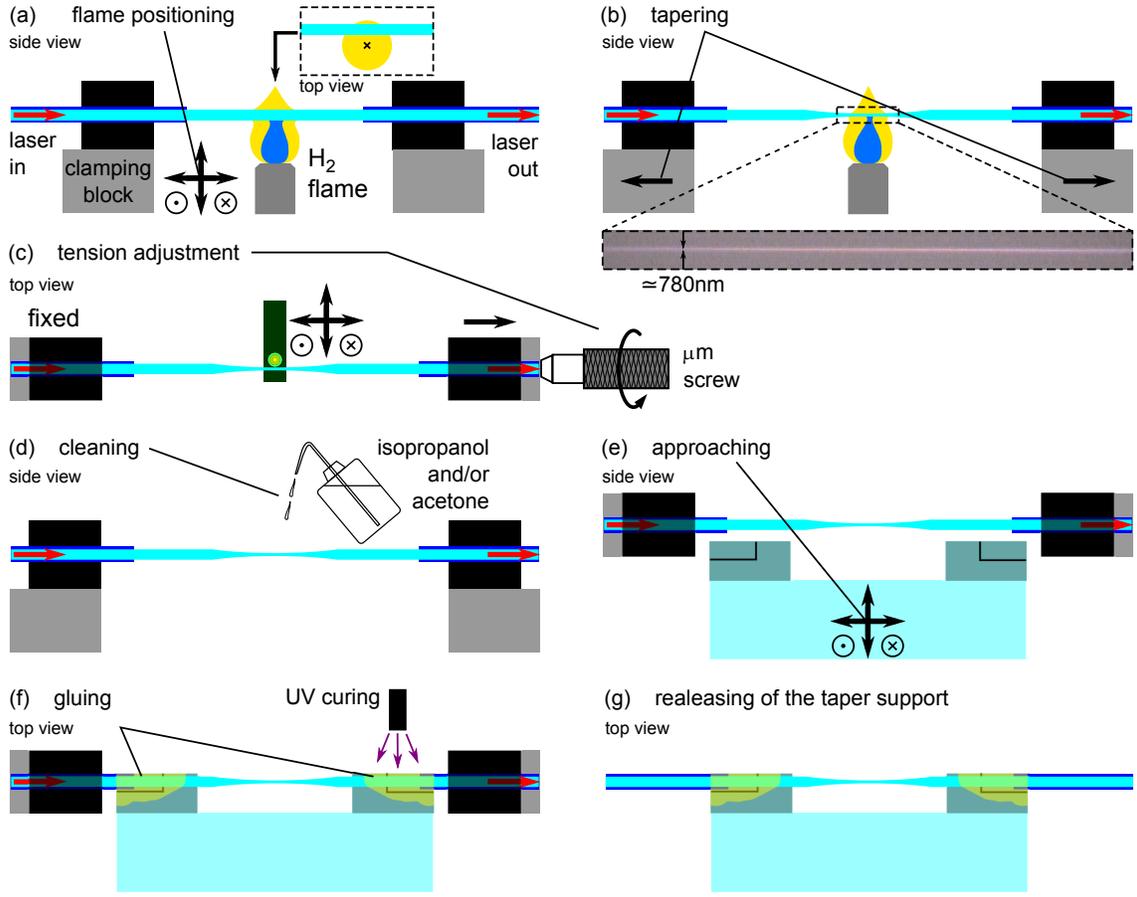}
  \caption{Fabrication and installation steps of the cryotaper. The inset shows a micrograph of a single-mode tapered-fiber fabricated here to transmit only a fundamental mode at 780\,nm wavelength.}
  \label{figure5}
\end{figure*}

This novel construction requires a special fabrication technique (Fig.~\ref{figure5}) adapted from the usual fiber taper fa\-brication protocol (see Ref.~\cite{Ding2010} for a complete description although the glass is melted here with an electric heater).
The tapered fiber is obtained by melting a standard SMF-28 silica optical fiber with a H$_2$ flame while stretching it until the melted region reaches a diameter of the order of the wavelength.
The tapered fiber supports then single mode waveguide propagation in the glass rod, where the surrounding air constitutes the cladding medium.
By adjusting the flame at a height as described in Fig.~\ref{figure5} (a), high transmission ($> 95\%$) tapers are typically obtained.
Pulling the flame further or closer to the fiber allows to define the glass fiber length subjected to the H$_2$ flame and therefore the final length of the tapered region.
Taking advantage of this, the tapered fiber is fabricated such as to ensure gluing of the remaining acrylate buffer to the C-shaped glass plate, improving the mechanical robustness of the cryotaper.
In Fig.~\ref{figure5} (c), the tension of the taper is tested by approaching an auxiliary toroid, snapping it to the taper, retracting it and measuring the ``snap-back distance'' \cite{Schliesser2010}.
Until this reaches less than a typical toroid radius, the taper is tensioned further using the micrometer screw displacing the fiber holder.
Cleaning, using droplets of liquid acetone or isopropanol, can eventually safely be performed to remove dusts that may aggregate during the previous steps and scatter off light, reducing the overall transmission.
Typically, however, the entire operation leaves the fiber transmission unchanged to $> 95$ \%.
After approaching and gluing under microscope monitoring (Fig. \ref{figure5} (e) and (f)), the cryotaper is released (Fig. \ref{figure5} (g)) and placed on the cryohead.
The taper is fabricated using one long optical fiber to avoid having fragile fiber splices inside the cooled experimental chamber.
The overall insertion loss is then typically below 20\%, mainly due to the fiber bending imposed by the confined experimental chamber.

\section{Measurement of the thermalization to the cryostat's base temperature}

\begin{figure}[t]
  \centering
  \includegraphics[scale=0.85]{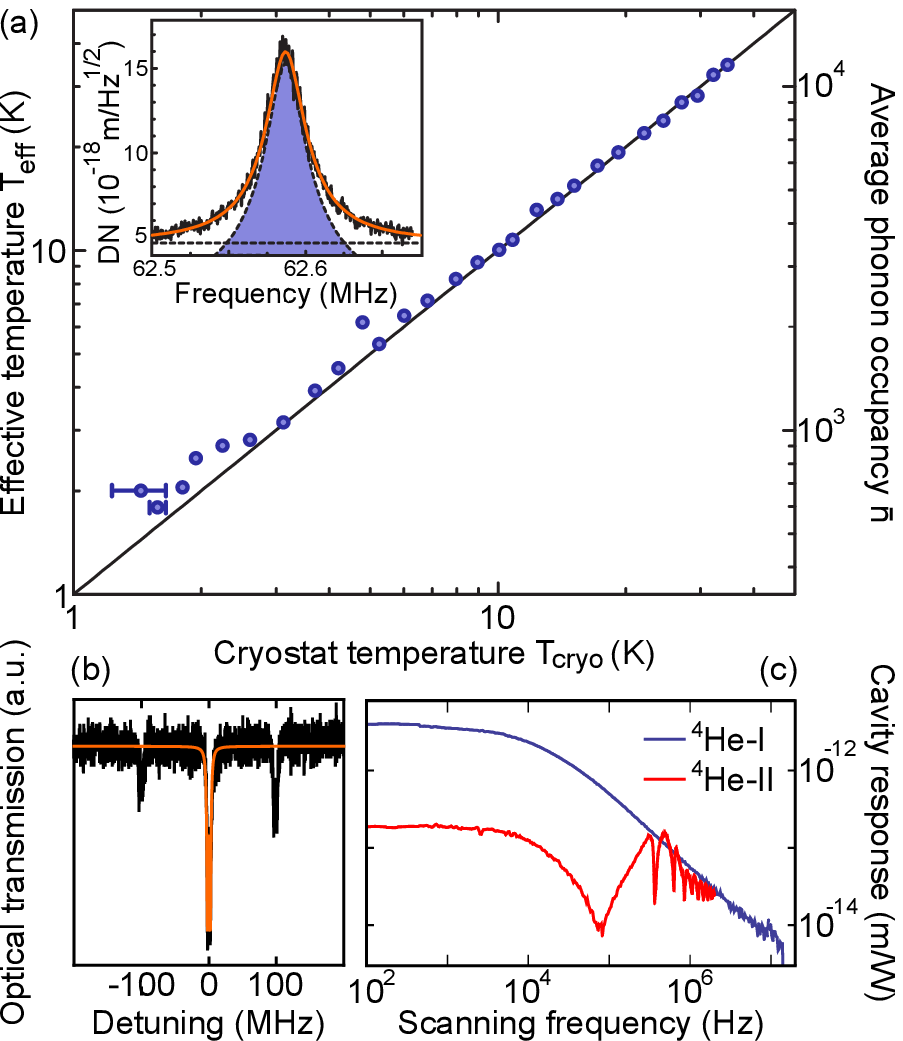}
  \caption{Thermalization in the $^4$He cryostat.
(a) Effective temperature of the mechanical mode versus the cryostat's temperature measured with the commercial Si diode. 
The correspondence with the guide to the eye proves the therma\-lization of the mode.
Inset: Mechanical displacement noise (DN) spectrum taken at $1.65$\,K.
The red line is a fit of the mechanical spectrum with the background, both separately represented by the dashed lines.
The effective temperature of the mechanical mode is extracted from the mechanical trace.
(b) Split optical resonance with a loaded optical Q of $\sim 0.8 \times 10^8$ taken at $\sim2$\,K with calibration sidebands.
(c) Amplitude spectrum of the thermal response of the toroid covered with a layer of $^4$He-I and of $^4$He-II, across the phase transition around $2.2$\,K and $\sim50$\,mbar of surrounding gaseous He.
}
  \label{figure6}
\end{figure}

The experimental setup described in the previous sections enables us to probe the optical whispering-gallery modes of a cavity placed in cryostat (Fig.~\ref{figure6} (b)).
In the context of optomechanics, the undertaken efforts are in vain if the mechanical mode of the sample does not thermalize to the base temperature of the cryostat.
Ideally, the refrigeration mechanism has to extract all the heat released by all sources, including the absorption of the high circulating power.
Here, we restrict our study to the cryogenic thermalization of the optomechanical system probed by a very weak laser.
By measuring an excellent thermalization down to 1.65\,K, we prove that the technology based on exchange gas enables optimal coo\-ling and that all other sources of heat, except intracavity power absorption, do not influence the temperature of the mechanical mode.

To perform this measurement, we take advantage of the optical coupling capabilities of the setup to record the thermally-driven, mechanically-induced frequency fluctuations with a Pound-Drever-Hall (PDH) detection technique \cite{Drever1983,Black2001,Arcizet2006c}.
The cavity (frequency $\omega_\mathrm{c}$) is resonantly probed by a weak laser ($1\,\mu$W input power), phase-modulated at a frequency $\Omega_\mathrm{PDH}$ exceeding the linewidth $\kappa$ of the cavity.
The photodetection signal of the optical cavity transmission carries then a part oscillating at $\Omega_\mathrm{PDH}$, corresponding to the beat of the carrier and the modulation sidebands.
The DC part of the demodulation of this voltage provides the error signal (switching sign at resonance) and is fed back into the frequency control of the laser.
The AC part, however, is spectrally analyzed with a resolution bandwidth $\Delta \omega_\mathrm{RBW}$ to record the mechanically induced frequency fluctuations of variance
\begin{equation}
  \langle \delta \omega_{\mathrm{c}}^{2} \rangle =  G^{2}  \langle \delta x^2 \rangle,
\end{equation}
where $G = - \omega_\mathrm{c} / R$ is the optomechanical coupling constant and $x$ the oscillator's displacement coordinate \cite{Arcizet2009}.
Note that the cavity radius $R$ can be correctly estimated using an optical microscope.
To calibrate the measured spectrum (Fig.~\ref{figure6} (a), inset), it is compared to a spectrally adjacent calibration peak \cite{Gorodetsky2010}.
This peak is generated from the coherent phase modulation of the probe beam at a frequency $\Omega_\mathrm{cal}$ very close to the mechanical frequency $\Omega_\mathrm{m}$, giving a laser frequency variance
\begin{equation}
  \langle \delta \omega_{\mathrm{l}}^{2} \rangle = \beta_{\mathrm{cal}}^{2} \Omega_{\mathrm{cal}}^{2} / 2,
\end{equation}
where $\beta_{\mathrm{cal}}$ is the phase modulation depth.
The frequency difference $|\Omega_\mathrm{cal} - \Omega_\mathrm{m}|$ must be contained well within the linewidth of the cavity to ensure identical transduction of the coherent phase modulation and the mechanical motion but larger than several times the mechanical linewidth $\Gamma_\mathrm{m}$ to avoid coherent driving of the mechanical mode.

Figure \ref{figure6} (a) shows the effective temperature $T_\mathrm{eff}$ of a 63-MHz radial breathing mode versus the tempe\-rature of the exchange gas of the experimental chamber of the cryostat $T_\mathrm{cryo}$, measured with the Si diode.
$T_\mathrm{eff}$ is estimated from the Lorentzian integral of the thermally-driven mechanical spectrum proportional to $\langle \delta \omega_{\mathrm{c}}^{2} \rangle$ (Fig.~\ref{figure6}~(a), inset, blue area), according to the equipartition of the mean energy of the harmonic mechanical oscillator.
The excellent thermalization is clearly marked by the agreement of the experimental points to the linear dependence, from 40\,K down to 1.8\,K, the latter corresponding to an average phonon occupancy $\bar{n}$ of 600.
Effective temperatures down to 1.65\,K are eventually reached (cryostat temperature out of range of the Si diode, data not shown in Fig.~\ref{figure6} (a)).
In this temperature range, the study in Ref.~\cite{Arcizet2009} not taking into conside\-ration the absorption mechanisms estimates a heating per intracavity power of 8.6\,K/W at an environmental pressure of $\sim0.5$\,mbar.
The good temperature control is additionally proven in \ref{figure6} (c) by recording the $^4$He-I to $^4$He-II phase transition inducing a drastic thermal res\-ponse spectrum change, measured here using a pump-and-probe scanning setup \cite{Riviere2010}.
In this regime, it can be evidenced that the thermal response measurement exhibits a series of resonances spaced by ca.\ 200\,kHz that may be a manifestation of a specific oscillatory mechanism in the superfluid part of the film.

\section{Conclusion}

We have introduced an experimental setup enabling the optical probing and the refrigeration of a silica whispering-gallery microtoroidal resonator.
By adap\-ting a straight cryotaper on a dedicated cryoprobe, we can optically couple the cavity with state-of-the-art efficiency and stability.
Furthermore, the viability of the approach is proven by measuring the thermalization of the mechanical mode down to the base temperature of 1.65\,K of the exchange-gas cryostat, corresponding to an ave\-rage phonon occupancy of less than 600.
Combined with laser cooling, one order of magnitude lower occupancy is even achieved with this setup, a performance then only limited by the heating induced by the absorption of the required high intracavity power \cite{Schliesser2009a}.

The proposed solutions are versatile: they allow tapered-fiber coupling at cryogenic temperatures of microcavities of various sizes and shapes, including polished crystalline resonators \cite{Hofer2010} and silica microspheres \cite{Takashima2010}, as well as any other resonator that can be coupled with straight tapered fiber such as photonic crystals on mesa structures \cite{Gavartin2011}.
Moreover, the cooling principle can be extended to $^3$He exchange-gas cooling in optomechanics experiments approaching the motional ground state \cite{Riviere2011} and the quantum-coherent coupling \cite{Verhagen2011a}.

Apart from its key role as a technical solution for those experiments, by its specific configuration the developed setup enables additionally the measurements of optical and mechanical properties of silica microcavities at low temperatures \cite{Arcizet2009} and of the intriguing aspects of superfluid helium-4, optically detected in such structures \cite{Riviere2010}.
The versatility of the tool and the outstanding measurement sensitivity provided by the cavity allows even to envisage complementary studies on the interaction of superfluid layers with a wide variety of materials \cite{Luhman2006}.

\begin{acknowledgments}
T.~J.~K.\ acknowledges funding by an European Research Council Starting Grant and the Deutsche Forschungsgemeinschaft.
R.~R.\ thanks Pierre Thoumany and Thomas Becker for the useful advice about cryo\-genics, and Thibaut Karassouloff, Samuel Del\'{e}glise and Antoine Heidmann for the assistance in characterizing the vibrational properties of the cryostat.
The Max-Planck-Institut f\"{u}r Quantenoptik is acknowledged for hosting this experiment.
\end{acknowledgments}


\end{document}